\documentclass[aps,prd,twocolumn,floatfix,showpacs]{revtex4} 

\usepackage{graphics}
\usepackage{graphicx}

\usepackage{latexsym}
\usepackage[cp866]{inputenc}
\usepackage[english]{babel}

\input epsf

\begin{document}

\title{\boldmath Description of the
$\psi(3770)$ resonance interfering with the background}
\author{N.N. Achasov\footnote{achasov@math.nsc.ru}
and G.N. Shestakov\footnote{shestako@math.nsc.ru}}
\affiliation{Laboratory of Theoretical Physics, S.L. Sobolev
Institute for Mathematics, 630090 Novosibirsk, Russia}


\begin{abstract}
The parameters of the interfering $\psi(3770)$ resonance should be
determined from the data on the reactions $e^+e^-$\,$\to$\,$D\bar D$
with the use of the models satisfying the elastic unitarity
requirement. The selection of such models can be realized by
comparing their predictions with the relevant data on the shape of
the $\psi(3770)$ peak in the non-$D\bar D$ decay channels. Here, we
illustrate this unitarity approach by the example of the most simple
variant of the model of the mixed $\psi(3770)$ and $\psi(2S)$
resonances. When new high-statistics data become available, it will
be interesting to test this clarity variant.
\end{abstract}


\maketitle

In the recent paper \cite{AS12}, we considered a few unitarized
models available for phenomenological description of the
$e^+e^-$\,$\to $\,$D\bar D$ reaction cross section in the
$\psi(3770)$ resonance region. Such models allow us to avoid the
spurious ambiguities in the interfering $\psi(3770)$ resonance
parameters determination, which have been recently revealed by
experimentalists when using unitarily uncorrected parametrizations
\cite{Bu,To2,An2,Li1,PDG12}.

In this report we present the simplest working variant of the model
of the mixed $\psi(3770)$ and $\psi(2S)$ resonances for the
description of interference phenomena in the $\psi(3770)$ region. It
was not discussed in Ref. \cite{AS12}. Owing to own clarity and
simplicity this variant can be tested, in the first place, in the
treatment of new high-statistics data which can be expected from
CLEO-c and BESIII \cite{Li1,Li,Hu,As09,RZC10,Br11} on the
$\psi(3770)$ shape in $e^+e^-$\,$\to $\,$D\bar D$. Here we also
concentrate great attention on the possibility of testing
theoretical models by comparing their predictions with the relevant
data on the shape of the $\psi(3770)$ peak in the non-$D\bar D$
decay channels, which are also expected from BESIII
\cite{As09,RZC10,Br11}.

In constructing the model describing the process $e^+e^-$\,$\to$\,$D
\bar D$, one must keep in mind that we investigate above all the
$D$-meson isoscalar electromagnetic form factor $F^0_D$. The phase
of $F^0_D$ in the elastic region [i.e., between the $D\bar D$
($\approx3.739$ GeV) and $D\bar D^*$ ($\approx3.872 $ GeV)
thresholds] is fixed by the unitarity condition equal to the phase
$\delta^0_1$ of the strong $P$-wave $D\bar D$ scattering amplitude
$T^0_1$ in the channel with isospin $I$\,=\,0, i.e.,
\begin{equation}\label{FD0}
F^0_D=e^{i\delta^0_1}\mathcal{F}^0_D \,,\end{equation} where
$\mathcal{F}^0_D$ and $\delta^0_1$ are the real functions of energy.
A similar representation of the amplitude $e^+e^-$\,$\to$\,$D\bar D$
used for the data description guarantees the unitarity requirement
on the model level \cite{AS12}. The sum of the
$e^+e^-$\,$\to$\,$D\bar D$ reaction cross sections is given by
\begin{equation}\label{CSeeDD} \sigma(e^+e^-\to D\bar D)=
\frac{8\pi\alpha^2}{3s^2}\left|F^0_D(s)\right|^2 \nu(s)\,,
\end{equation} where $s$ is the $D\bar D$-pair invariant mass square,
$\nu(s)=[p^3_0(s)+p^3_+(s)]/\sqrt{s}$, $\,p_{0,+}(s)$\,=\,$\sqrt
{s/4- m^2_{D^{0,+}}}$ and $\alpha$\,=\,$e^2/4\pi$\,=\,1/137 (here we
do not touch on the questions about the isospin symmetry breaking).
Below, for short $\psi(3770)$ is denoted as $\psi''$.

\begin{figure}\centerline{\epsfysize=3.3in
\epsfbox{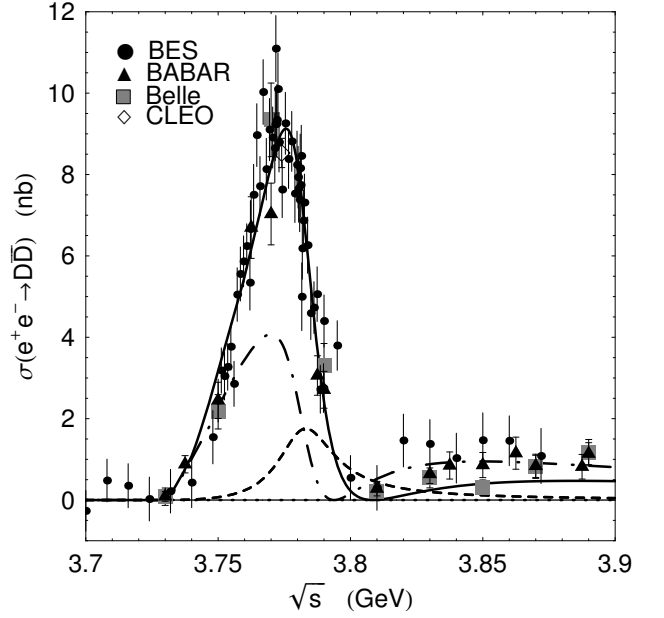}} 
\caption{The simplest variant of the model of the mixed $\psi''$ and
$\psi(2S)$ resonances. The solid curve is the fit using Eqs.
(\ref{CSeeDD})--(\ref{RDDMix1}) to the data from BES
\cite{Ab3a,Ab4}, CLEO \cite{Be2}, $BABAR$ \cite{Au1,Au2}, and Belle
\cite{Pa} for $\sigma(e^+ e^-$\,$\to$\,$D\bar D)$. The dashed and
dot-dashed curves show the contributions to the cross section from
the $\psi''$ and $\psi(2S)$ production amplitudes proportional to
the products of the coupling constants $g_{\psi''\gamma}g_{\psi''
D\bar D}$ and $g_{\psi(2S)\gamma}g_{ \psi(2S)D\bar D}$,
respectively; see Eqs. (\ref{F0DMix}) and (\ref{RDDMix1}). For more
details on the data see Ref. \cite{AS12}.} \label{Figure1}
\end{figure}

Consider now the model which takes into account in $F^0_D$ and
$T^0_1$ the contributions only from the $\psi''$ and $\psi(2S)$
resonances. Owing to the common $D^0\bar D^0$ and $D^+D^-$ coupled
channels, the $\psi''$ and the $\psi(2S)$ can transform into each
other (i.e., mix); for example, $\psi''$\,$\to$\,$D\bar
D$\,$\to$\,$\psi (2S)$. The form factor $F^0_D$, corresponding to
the contribution of the mixed $\psi''$ and $\psi(2S)$ resonances,
can be represented in the following symmetric form
\cite{AS12,A1,A2,A3}:
\begin{equation}\label{F0DMix}F^0_D(s)=\frac{
\mathcal{R}_{D\bar D}(s)}{D_{\psi''}(s)D_{\psi (2S)}(s)-\Pi^2_{
\psi''\psi(2S)}(s)}\,,\end{equation} where $D_{\psi''}(s)$ and
$D_{\psi(2S)}(s)$ are the inverse propagators of $\psi''$ and
$\psi(2S)$, respectively,
\begin{equation}\label{Dpsi2}
D_{\psi''}(s)=m^2_{\psi''}-s-i\sqrt{s}\Gamma_{\psi''D\bar D}(s)\,,
\end{equation}
\begin{equation}\label{Dpsi1}
D_{\psi(2S)}(s)=m^2_{\psi(2S)}-s-i\sqrt{s}\Gamma_{\psi(2S)D\bar
D}(s)\,,
\end{equation}
\begin{equation}\label{Gpsi2DD}
\Gamma_{\psi''D\bar D}(s)=\frac{g^2_{\psi''D\bar
D}}{6\pi}\,\frac{\nu(s)}{\sqrt{s}}\,,
\end{equation}
\begin{equation}\label{Gpsi1DD}
\Gamma_{\psi(2S)D\bar D}(s)=\frac{g^2_{\psi(2S)D\bar
D}}{6\pi}\,\frac{\nu(s)}{\sqrt{s}}\,,
\end{equation}
\begin{eqnarray}\label{RDDMix}
& \mathcal{R}_{D\bar D}(s) & \nonumber\\ & =g_{\psi(2S)
\gamma}[D_{\psi''}(s)g_{\psi(2S)D\bar D}+
\Pi_{\psi''\psi(2S)}(s)g_{\psi''D\bar D}] &  \nonumber\\
& +g_{\psi''\gamma}[D_{\psi(2S)}(s)g_{\psi''D\bar D}+
\Pi_{\psi''\psi(2S)}(s)g_{\psi(2S)D\bar D}]. & \end{eqnarray} The
constants $g_{\psi''D\bar D}$, $\,g_{\psi(2S)D\bar D}$, and
$\,g_{\psi''\gamma}$, $\,g_{\psi(2S)\gamma}$ characterize couplings
of the $\psi''$, $\psi(2S)$ to the $D\bar D$ and virtual $\gamma$
quantum, respectively. The amplitude $\Pi_{\psi''\psi(2S)}(s)$
describing the $\psi''-\psi(2S)$ mixing has the form
\begin{equation}\label{Ppsi2psi1}
\Pi_{\psi''\psi(2S)}(s)=\mbox{Re}\Pi_{\psi''\psi(2S)}(s)+i\,\frac{
g_{\psi''D\bar D}g_{\psi(2S)D\bar D}}{6\pi}\,\nu(s).
\end{equation} Its imaginary part is due to the $\psi''$\,$\to$\,$ D\bar
D$\,$\to$\,$\psi(2S)$ transitions via the real $D\bar D$
intermediate states. Substituting Eqs.
(\ref{Dpsi2})--(\ref{Gpsi1DD}) and (\ref{Ppsi2psi1}) into Eq.
(\ref{RDDMix}), it is easy to make certain that $\mathcal{R}_{D\bar
D}(s)$ is a real function. Thus, the model can explain the dip
observed in $\sigma(e^+e^-$\,$\to $\,$D\bar D)$ near
$\sqrt{s}\approx3.81$ GeV (see Fig. \ref{Figure1}) by the zero in
$F^0_D(s)$, caused by compensation between the $\psi''$ and
$\psi(2S)$ contributions. Note that $\mbox{Re}\Pi_{\psi''\psi(2S)}
(s)$ cannot be strictly calculated. Its approximations, for example,
by the expression $c_0+sc_1$, where $c_0$ and $c_1$ are free
parameters, can be used as a resource for the fit improvement.
Below, for simplicity we put
$\mbox{Re}\Pi_{\psi''\psi(2S)}(s)$\,=\,0. Then Eq. (\ref{RDDMix})
takes the form
\begin{eqnarray}\label{RDDMix1} & \mathcal{R}_{D\bar
D}(s)=(m^2_{\psi''}-s)g_{\psi(2S)\gamma}g_{\psi(2S)D\bar D}
& \nonumber\\
& +(m^2_{\psi(2S)}-s)g_{\psi''\gamma}g_{\psi''D\bar D}\,.&
\end{eqnarray}

The curves in Fig. \ref{Figure1} correspond to
$m_{\psi''}$\,=\,3.794 GeV, $g_{\psi''D\bar D}=\pm$\,14.35 [i.e.,
$\Gamma_{\psi''D\bar D}(m^2_{\psi''})\approx56.8$ MeV, see Eq.
(\ref{Gpsi2DD})], $g_{\psi''\gamma}=\pm$\,0.1234 GeV$^2$ [i.e.,
$\Gamma_{\psi''e^+e^-}=4\pi\alpha^2
g^2_{\psi''\gamma}/(3m^3_{\psi''})\approx0.062$ keV], and
$g_{\psi(2S)D\bar D}=\pm$\,$20.11$. In so doing, if
$g_{\psi''\gamma}g_{\psi''D\bar D}>0$ ($<0$), then
$g_{\psi(2S)\gamma}g_{\psi(2S)D\bar D}<0$ ($>0$), see Eq.
(\ref{RDDMix1}). The values $m_{\psi(2S)}=3.6861$ GeV and
$g_{\psi(2S) \gamma}=\pm0.7262$ GeV$^2$ were fixed according the
data \cite{PDG12} and the relation $\Gamma_{\psi(2S)e^+e^-}=4\pi
\alpha^2 g^2_{\psi(2S)\gamma}/(3m^3_{\psi(2S)})=2.35$ keV.

The values of the fitted parameters $\,m_{\psi''}$,
$\,g_{\psi''D\bar D}$, and $\,g_{\psi''\gamma}$ can essentially
depend on the model used for the description of the total
contribution of the $\psi''$ resonance  and background. The analysis
\cite{AS12} indicates that the components of the $e^+e^-\to D\bar D$
amplitude can be very different in the different models. For the
model of the mixed $\psi''$ and $\psi(2S)$ resonances, the
contributions of the components in question are shown in Fig.
\ref{Figure1} by the dashed and dot-dashed curves. On the other
hand, it is clear that the interference pattern in the $\psi''$
region depends on the reaction. Therefore, the selection of the
theoretical models should be carry out by comparing their
predictions with the experimental data on the shape of the $\psi''$
peak for several different reactions.

\begin{figure}\centerline{\epsfysize=3.0in 
\epsfbox{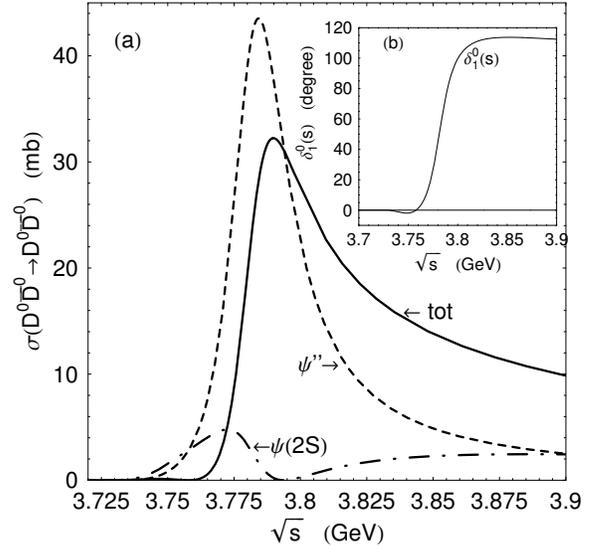}} 
\caption{The predictions of the model with the mixed $\psi''$ and
$\psi(2S)$ resonances. (a) The solid, dashed, and dot-dashed curves
correspond to $\sigma(D^0\bar D^0\to D^0\bar D^0)$\,=\,$3\pi
|\sin\delta^0_{1}(s)|^2/p^2_0(s)$ and the $\psi''$ and $\psi(2S)$
contributions proportional to $g^2_{\psi''D\bar D}$ and $g^2_{
\psi(2S)D\bar D}$ in Eq. (\ref{T10}), respectively. (b) The phase
$\delta^0_{1}(s)$.} \label{Figure2}
\end{figure}

For example, after the fitting of the $e^+e^-\to D\bar D$ data we
all know about $D\bar D$ elastic scattering in the $P$-wave at the
model level,
\begin{eqnarray}\label{T10}
T^0_1(s)=e^{i\delta^0_1(s)}\sin\delta^0_1(s)=\frac{\nu(s)}{6\pi}
\qquad\qquad & \nonumber\\
\times\left[\frac{(m^2_{\psi''}-s)g^2_{\psi (2S)D\bar
D}+(m^2_{\psi(2S)}-s)g^2_{\psi''D\bar D}}{D_{\psi''}(s)
D_{\psi(2S)}(s)-\Pi^2_{\psi''\psi(2S)}(s)}\right].&&
\end{eqnarray}
The corresponding cross section and phase are shown in Fig.
\ref{Figure2}. Unfortunately, these predictions are not possible to
verify. However, there are other processes which can be measured
experimentally.

We are interested in the interference phenomena in the $\psi''$
region in the reactions $e^+e^-$\,$\to$\,non-$D\bar D$. We confine
ourselves to the simplest non-$D\bar D$ final states, the form
factors of which are determined by a single independent invariant
amplitude. Such reactions are $e^+e^-$\,$\to$\,$\gamma \chi_{c0}$,
$\,\gamma\eta_c$, $\,\gamma\eta'$, $\,J/\psi\eta$, $\,\phi\eta\,$,
and so on.

\begin{table*} 
\caption{Information about the $\psi(2S)$ \cite{PDG12} and $\psi''$
\cite{PDG12,Br,Ped,Am,As} resonances in non-$D\bar D$ decay channels
(ab).} \vspace{0.05cm}
\begin{tabular}{|c|c|c|c|c|c|c|}
\hline $ab$ & $B(\psi(2S)\to ab)$ & $\Gamma_{\psi(2S)ab}$ (keV) & $g_{\psi(2S)ab}$ (GeV$^{-1}$) & $B(\psi''\to ab)$ & $\Gamma_{\psi''ab}$ (keV) &
$\sigma(e^+e^-$\,$\to$\,$ab$) (pb)\\
\hline $\gamma\chi_{c0}$ & $(9.68\pm0.31)\%$ & $29.4\pm0.9$ & $\pm(0.250\pm0.004)$ & $(7.3\pm0.9)\times10^{-3}$ & $172\pm30$ & $72\pm9$ \\
\hline $\gamma\eta_c$    & $(3.4\pm0.5)\times10^{-3}$ & $1.03\pm0.15$ & $\pm(1.22\pm0.09)\times10^{-2}$    & $\cdot\cdot\cdot$ & $\cdot\cdot\cdot$ & $\cdot\cdot\cdot$ \\
\hline $\gamma\eta'$     & $(1.23\pm0.06)\times10^{-4}$ & $0.374\pm0.018$ & $\pm(1.67\pm0.04)\times10^{-3}$ & $<1.8\times10^{-4}$   & $\cdot\cdot\cdot$ & $\cdot\cdot\cdot$ \\
\hline $J\psi\eta$       & $(3.28\pm0.07)\%$ & $9.97\pm0.21$   & $\pm(0.218\pm0.002)$ & $(9\pm4)\times10^{-4}$ & $21\pm10$ & $5.5\pm2.5$ \\
\hline $\phi\eta$ & $(2.8^{+1.0}_{-0.8})\times10^{-5}$ & $(8.5^{+3.0}_{-2.4})\times10^{-3}$ & $\pm(2.7^{+0.5}_{-0.4})\times10^{-4}$ & $(3.1\pm0.7)
\times10^{-4}$ & $7.4\pm1.6$ & $2.4\pm0.6$ \\
\hline\end{tabular}\end{table*}   

The cross section for $e^+e^-$\,$\to$\,$ab$
($ab$\,=\,$\gamma\chi_{c0}$, $\gamma\eta_c$, $\gamma\eta'$,
$J/\psi\eta$, $\phi\eta$) in the $\psi''$ region can be written as
\begin{equation}\label{SigmAB} \sigma(e^+e^-\to ab)=\frac{4\pi\alpha^2
k^3_{ab}(s)} {3s^{3/2}}\left|F_{ab}(s)\right|^2\,,\end{equation}
where
$k_{ab}(s)$\,=\,$\sqrt{[s-(m_a+m_b)^2][s-(m_a-m_b)^2]}\,/(2\sqrt{s})$
and $F_{ab}(s)$ is the electromagnetic form factor of the $ab$
system. Equation (\ref{SigmAB}) implies that the decay amplitude of
the virtual timelike photon with the mass $\sqrt{s}$ into
$\gamma\chi_{c0}$ ($\chi_{c0}$ is the scalar meson) is given by
\begin{equation}\label{V1}
eF_{\gamma\chi_{c0}}(s)\,\epsilon^\gamma_\mu(q)\epsilon^\gamma_\nu
(k)(q\cdot k\,g_{\mu\nu}-k_\mu q_\nu)\,,\end{equation} where
$\epsilon^\gamma_\mu(q)$ and $\epsilon^\gamma_\nu (k)$ are the
polarization four-vectors of the intermediate (virtual) and final
photons with four-momenta $q$ ($q^2$\,=\,$s$) and $k$, respectively;
and, its decay amplitude into $V0^-$ ($0^-$ denotes a pseudoscalar
meson and $V0^-$\,=\, $\gamma\eta_c$, $\gamma\eta'$, $J/\psi\eta$,
$\phi\eta$) is given by
\begin{equation}\label{V2}eF_{V0^-}(s)\,\varepsilon_{
\mu\nu\sigma\tau}\epsilon^\gamma_\mu(q)\epsilon^V_\nu(k)q_\sigma
k_\tau\,.\end{equation} In the model under consideration we may
write \begin{equation}\label{FAB}
F_{ab}(s)=\frac{\mathcal{R}_{ab}(s)}{D_{\psi''}(s)D_{\psi
(2S)}(s)-\Pi^2_{\psi''\psi(2S)}(s)}\,,
\end{equation} where
\begin{eqnarray}\label{RAB} & \mathcal{R}_{ab}(s) & \nonumber\\
&=g_{\psi(2S)\gamma}[D_{\psi''}(s)g_{
\psi(2S)ab}+\Pi_{\psi''\psi(2S)}(s)g_{\psi''ab}] & \nonumber\\
& +g_{\psi''\gamma}[D_{\psi(2S)(s)}g_{\psi''ab}+ \Pi_{\psi''\psi
(2S)}(s)g_{\psi(2S)ab}]&\end{eqnarray} and $g_{\psi(2S)ab}$,
$g_{\psi''ab}$ are the effective coupling constants of the $\psi
(2S)$, $\psi''$ to the $ab$ channel. These coupling constants are
taken into account in $F_{ab}(s)$ in the first order of perturbation
theory. Their relative smallness is caused by the electromagnetic
interaction for the $\gamma\chi_{c0}$ and $\gamma\eta_c $ channels,
by the dynamics of the Okubo-Zweig-Iizuka rule violation \cite{AK1,
AK2,AK3} for the $J/\psi\eta$ and $\phi\eta$ channels, and by a
combination of the above reasons for the $\gamma\eta'$ channel.

As a first (rough) approximation, we suppose that the coupling
constants for radiative transitions between charmonium states
$(c\bar c)_i\to\gamma\,(c\bar c)_f$ [index $i$ ($f$) labels initial
(final) state] and also those for hadronic transitions $(c\bar
c)_i\to(c\bar c)_f\,h$ and radiative decays $(c\bar
c)_i\to\gamma\,h$, probing the gluon content of light hadrons $h$,
are real \cite{As09,Br11,KS}. That is, we neglect the contributions
of the real $D\bar D$ intermediate states, taking into account which
leads to the appearance of imaginary parts of effective coupling
constants \cite{AK1,AK2,AK3}. High-statistics studies of the
$e^+e^-$\,$\to$\,non-$ D\bar D$ processes in the $\psi''$ region
will show how this is justified. Note that for the $(c\bar
c)_i\to\phi\eta$ decay the $D\bar D$ loop rescattering mechanism
$(c\bar c)_i\to D\bar D\to\phi\eta$ is suppressed by the
Okubo-Zweig-Iizuka rule. The phase of the $\phi\eta$ final state
interaction is unknown. However, this phase is common for different
contributions to $e^+e^-\to\phi\eta$ and does not appear in the
cross section. At this stage, we do not take into account the
interference between the $e^+e^-\to(c\bar c)\to\phi\eta$ amplitude
and the background from the light quark production $e^+e^-\to(s\bar
s)\to\phi\eta$. With the above assumptions, the effective coupling
constants $g_{\psi(2S)\phi\eta}$ and $g_{\psi''\phi\eta}$ will be
considered to be real as well.

Table I presents information about the $\psi(2S)$ \cite{PDG12} and
$\psi''$ \cite{PDG12,Br,Ped,Am,As} resonances in the $ab$ decay
channels, which we use to construct the corresponding mass spectra.
The values for $g_{\psi(2S)ab}$ indicated in the table are obtained,
up to the sign, from the data on the $\psi(2S)$\,$\to$\,$ab$ decay
widths by the formula
\begin{equation}\label{Gpsi2SAB}\Gamma_{\psi(2S)ab}=\frac{g_{\psi(2S)ab}^2}
{12\pi}k^3_{ab}(m^2_{\psi(2S)})\,,\end{equation} which implies that
the amplitudes of the $\psi(2S)\to\gamma\chi_{c0}$ and $\psi(2S)\to
V0^-$ decays have the form
\begin{eqnarray}\label{V3V4}
& g_{\psi(2S)\gamma\chi_{c0}}\,\epsilon^{\psi(2S)}_\mu(q)
\epsilon^\gamma_\nu(k)(q\cdot k\,g_{\mu\nu}-k_\mu q_\nu) & \nonumber \\
& \mbox{and}\ \ g_{\psi(2S)V0^-}\,\varepsilon_{
\mu\nu\sigma\tau}\epsilon^{\psi(2S)}_\mu(q)\epsilon^V_\nu(k)q_\sigma
k_\tau\,, & \nonumber\end{eqnarray} respectively. The relative signs
of the constants $g_{\psi(2S)ab}$ and $g_{\psi''ab}$ are unknown.
Therefore, the relative signs between the first and subsequent three
terms in Eq. (\ref{RAB}) (they are controlled by signs of the
coupling constant products) can be chosen in two ways: $(+-+)$ or
$(-+-)$. Here, we took into account the above-mentioned sign
correlation between $g_{\psi''\gamma}g_{ \psi''D\bar D}$ and
$g_{\psi(2S)\gamma}g_{\psi(2 S)D\bar D}$.

\begin{figure}\centerline{\epsfysize=5.20in 
\epsfbox{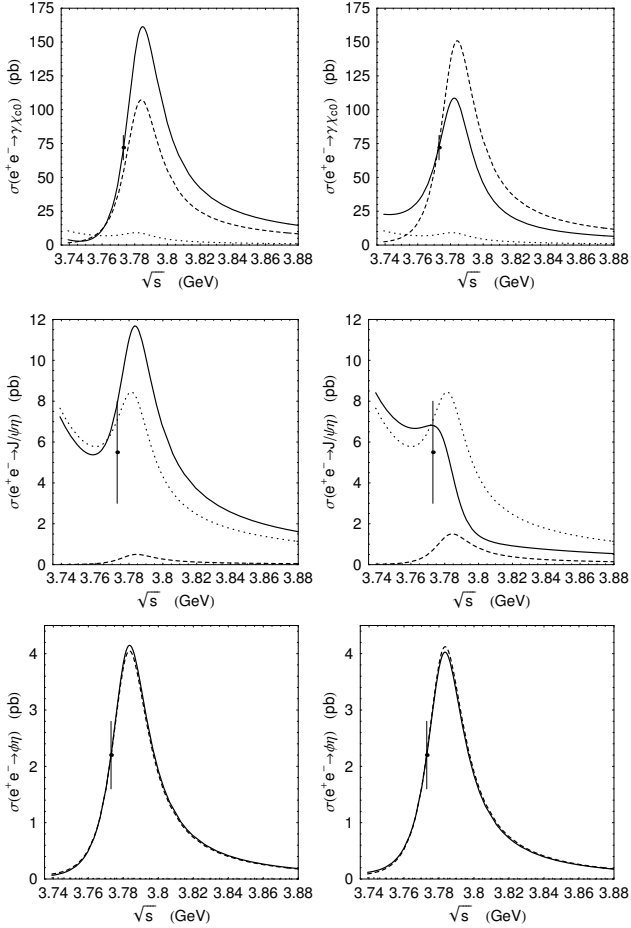}} 
\caption{The cross sections for $e^+e^-$\,$\to$\,$\gamma\chi_{c0}$,
$e^+e^-$\,$\to$\,$J/\psi\eta$, and $e^+e^-$\,$\to$\,$\phi\eta$
(left) for case $(+-+)$ and (right) for case $(-+-)$.}
\label{Figure3}
\end{figure}

The existing information about the
$\psi''$\,$\to$\,$\gamma\chi_{c0}$, $\gamma\eta_c$, $\gamma\eta'$,
$J/\psi\eta$, $\phi\eta$ decays are very poor. The CLEO
Collaboration measured the reactions $e^+e^-$\,$\to$\,$\gamma
\chi_{c0}$ \cite{Br}, $e^+e^-$\,$\to$\,$ J/\psi\eta$ \cite{Am}, and
$e^+e^-$\,$\to$\,$\phi\eta$ \cite{As} at a single point in energy
$\sqrt{s}=3773$ MeV (at the supposed maximum of cross sections). The
approximate values for $\sigma(e^+e^-$\,$\to$\,$ab$) are presented
in Table I and Fig. \ref{Figure3} by the points with the error bars.
They allow us to roughly estimate the coupling constants
$g_{\psi''\gamma\chi_{c0}}\approx\pm0.608$ GeV$^{-1}$,
$g_{\psi''J/\psi\eta}\approx\pm0.0375 $ GeV$^{-1}$,
$g_{\psi''\phi\eta}\approx\pm1.1\times10^{-2}$ GeV$^{-1}$ for case
$(+-+)$ and $g_{\psi''\gamma\chi_{c0}}\approx\pm0.721$ GeV$^{-1}$,
$g_{\psi''J/\psi\eta}\approx\pm0.065 $ GeV$^{-1}$,
$g_{\psi''\phi\eta}\approx\pm1.11\times10^{-2}$ GeV$^{-1}$ for case
$(-+-)$, by using Eqs. (\ref{SigmAB}), (\ref{FAB}), and (\ref{RAB}),
and construct the corresponding cross sections as functions of
energy.

The solid curves in Figs. \ref{Figure3} show the cross sections for
$e^+e^-$\,$\to$\,$\gamma\chi_{c0}$, $e^+e^-$\,$\to$\,$J/\psi\eta$,
and $e^+e^-$\,$\to$\,$\phi\eta$; the dashed and dotted curves show
the contributions from the $\psi''$ and $\psi(2S)$ resonances
proportional to [see Eq. (\ref{RAB})]
\begin{eqnarray}\label{RAB12}
&[g_{\psi''\gamma}D_{\psi(2S)}(s)+g_{\psi(2S)\gamma}\Pi_{\psi''
\psi(2S)}(s)]g_{\psi''ab} & \nonumber \\
&\ \ \mbox{and}\ \
[g_{\psi(2S)\gamma}D_{\psi''}(s)+g_{\psi''\gamma}\Pi_{\psi''
\psi(2S)}(s)]g_{\psi(2S)ab}\,,& \nonumber
\end{eqnarray}
respectively. The values of each of these contributions to
$F_{ab}(s)$ change from reaction to reaction according to changes of
$g_{\psi''ab}$ and $g_{\psi(2S)ab}$. At the same time, their
$s$-dependence does not change, as it has already been determined by
the model parameters found from fitting the $e^+e^-$\,$\to$\,$D\bar
D$ cross section (simultaneous fits to the data on the reactions
$e^+e^-$\,$\to$\,$D\bar D$ and $e^+e^-$\,$\to$\,non-$D\bar D$ is yet
to come). Note that the cross section for $e^+e^-$\,$\to$\,$\phi
\eta$ is completely dominated by the $\psi''$ contribution. Note
also that the Belle Collaboration has recently
\begin{figure}\centerline{\epsfysize=1.75in 
\epsfbox{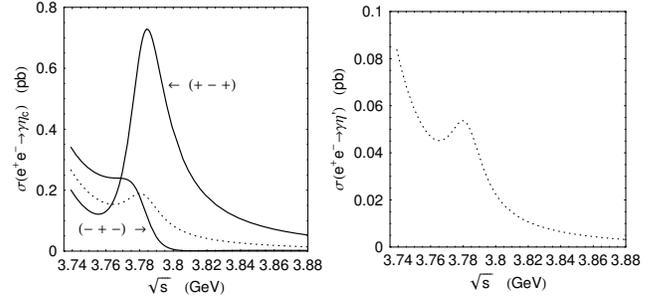}} 
\caption{The cross sections for $e^+e^-$\,$\to$\,$\gamma\eta_c$
(left) and $e^+e^-$\,$\to$\,$\gamma\eta'$ (right).} \label{Figure4}
\end{figure}
measured the cross section for $e^+e^-$\,$\to $\,$J/\psi\eta$
between $\sqrt{s}$\,=\,3.8 GeV and 5.3 GeV \cite{Wa}. Unfortunately,
the data for 3.8 GeV$\,<\sqrt{s}<\,$4 GeV have large errors, which
does not allow us to extract any useful information.

The cross sections for $e^+e^-$\,$\to $\,$\gamma\eta_c$ and
$e^+e^-$\,$\to$\,$\gamma \eta'$ in the $\psi''$ region are unknown.
Using information about the $\psi(2S)$ from Table I, we estimate the
cross sections at $g_{\psi''\gamma\eta_c}$\,=\,$g_{
\psi''\gamma\eta'}$\,=\,0. The results are shown in Fig.
\ref{Figure4} by the dotted curves. Here, as in the case of the
dotted curves in Fig. \ref{Figure3}, the resonant enhancement on the
tails of the $\psi(2S)$ contribution arises owing to the
$\psi''-\psi(2S)$ mixing. If we put $\Gamma_{\psi''\gamma\eta_c
}\approx1$ keV \cite{KS}, which corresponds to $g_{\psi''\gamma
\eta_c}\approx\pm1\times10^{-2}$ GeV$^{-1}$, then $\sigma(e^+e^-$\,$
\to$\,$\gamma\eta_c)$ takes the form shown in the left plot in Fig.
\ref{Figure4} by the solid curves for cases $(+-+)$ and $(-+-)$.

The above examples tell us that the mass spectra in the $\psi''$
region in the non-$D\bar D$ channels can be very diverse. Therefore,
we should expect that the data on such spectra, together with the
$e^+e^-$\,$\to $\,$D\bar D$ data, will impose severe restrictions on
the constructed dynamical models for the $\psi''$ resonance
interfering with the background.
\\[0.1cm]
\hspace*{0.25cm} This work was supported in part by RFBR, Grant No.
13-02-00039, and Interdisciplinary Project No. 102 of the Siberian
division of RAS.



\end{document}